\documentstyle[aps,preprint,epsf]{revtex}
\begin{document}
%\draft
\preprint{}
\title{  Enhancement of  coherent
       light amplification \\ using   Anderson localization  }
\author{ Prabhakar  Pradhan\cite{ppemail}}
\address{Department of Electrical Engineering\\
         University of California,  Los Angeles, CA 90095, USA}   
\maketitle
\date{\today}
\begin{abstract}
Several aspects of interplay between Anderson
localization and coherent amplification/absorption, and aspect
of mirrorless laser for a laser-active (amplifying) disordered
dielectric medium have been addressed. We have calculated the
statistics of the reflection coefficient 
and the associated phase for a light wave reflected
from a coherently amplifying/absorbing one-dimensional
disordered dielectric medium for different lengths, different
disorder strengths and different "amplification"/"absorption"
parameters of the disordered sample. Problems with modeling coherent
amplification/absorption by complex potential have been discussed. We
have shown an alternative way to model the coherent absorption.
Several conceptual and physical aspect for coherently
amplifying/absorbing media with disorder also have been discussed. 
\end{abstract}
\pacs{.}
%
%\vspace{18pt}
\section{introduction}
Wave propagation in disordered media and its consequences to the
disorder induced localization, or Anderson localization
\cite{anderson} have been studied in detail for electron (quantum)
waves \cite{lee}, as well as for light (classical) waves \cite{ping}
for last four decades. We have discussed several aspects of transport
and localization for 1D electronic systems in our previous paper
\cite{pp}. In this paper we are dealing with an interesting aspect
of the application of the Anderson localization to enhance  the coherent 
light amplification in a coherently amplifying disordered medium.
Due to the bosonic nature of the light quanta, there is possibility
of coherent amplification or coherent absorption(attenuation) for
light; and these possibilities are absent for electronic(fermionic)
case. Background to study this problem comes from the following
two facts: (1) coherent amplification in an coherently amplifying
medium is a non conserving scattering process where temporal phase
coherence of a wave is preserved despite amplification and (2)
coherent back scattering (CBS) in a disordered media is the main
cause of weak and strong localization will not be affected with
additional presence of a coherently amplifying medium due to the
persistence of phase coherence of the interfering waves. The
preservation of the CBS, hence the localization, despite
amplification can bring the interesting possibility of obtaining
synergitic enhancement of wave amplification , or laser action
without mirrors due to the confinement by the Anderson localization, in
an optically pumped laser active disordered media.
In our previous short paper \cite{ppnk} such possibility was
shown theoretically for 1D amplifying disordered media.
{\em Our calculation \cite{ppnk}  lead to an important 
result that in the presence
of amplification in a disordered optical medium, the Anderson
localization enhances the coherent amplification.}
In fact, recent experiments
\cite{lawandy1,ganack,lawandy2,albada,wths,prasad} 
support aspects of the lasing action, or more precisely, enhanced
Amplification of Spontaneous Emission (ASE), in an amplifying and
strongly scattering medium without an external resonant cavity. There
is current interest to see the affect of disorder on coherent
amplification/absorption in different disorder regimes.
 In this detailed paper
we have discussed the possible background and probable several
physical aspects of the problem of coherent light amplification using
the Anderson localization and a detailed study of the 1D
amplifying/absorbing disordered medium.

It is now well known that in 1D and 2D all states of a wave are
localized for both electronic and optical disordered systems
\cite{lee,ping}. This motivate one to study synergetic effect of
amplification and localization for lower dimensions. 
We study analytically and numerically, the probability
distribution of the reflection coefficient($|R|^2$) 
and phase ($\theta$) of the complex amplitude reflection 
coefficient ($R=|R|e^{i\theta}$) of a wave reflected from
a one-dimensional coherently absorbing/attenuating disordered
medium for a Gaussian white-noise potential for different
lengths of the sample, with different strengths of disorder, and
with different active (gain or loss) parameter values.
We set up a general framework to study 1D coherently
amplifying/absorbing disordered media and discuss the context
of physical realization for both optical and electronic cases.
Coherent amplification we mean mainly is due to the
stimulated emission of radiation which can only be applicable to an
optical (bosonic) case and has no electronic counterpart. Coherent
absorption can have physical situation in both the optical and the
electronic cases --- for the electronic(fermionic) case it signifies
effectively the incoherent part of the electron wave function
caused by inelastic scattering. We consider here mainly the 1D
Helmholtz equation, where the coherent (linear)
amplification/absorption is modeled by adding a constant
imaginary part to the real potential. We derive first a Langevin
equation for the complex amplitude reflection coefficient $R(L)$
and then the Fokker-Planck equation in the $(|R|^2,\theta)$-space
for evolution with the length of the sample for different
parameter values: (1) disorder strength, and (2) active
parameter. We discuss the appropriate physical situations 
for both the cases, i.e., absorption and amplification. 

 Modeling amplification/absorption by a complex potential,
however, always gives a reflection part due to the mismatch 
between the real and the imaginary potentials. We discuss these
issues in detail. We also derive a Langevin equation and then the FP
equation by a phenomenological modeling where absorption has no
concomitant reflection part. We also have discussed recent
experimental and theoretical works of others on coherent
amplification/absorption. 
\section{ Coherent backscattering in the presence of coherent
amplification/absorption in an active disordered medium } 
\begin{figure}
\centerline{\epsfbox{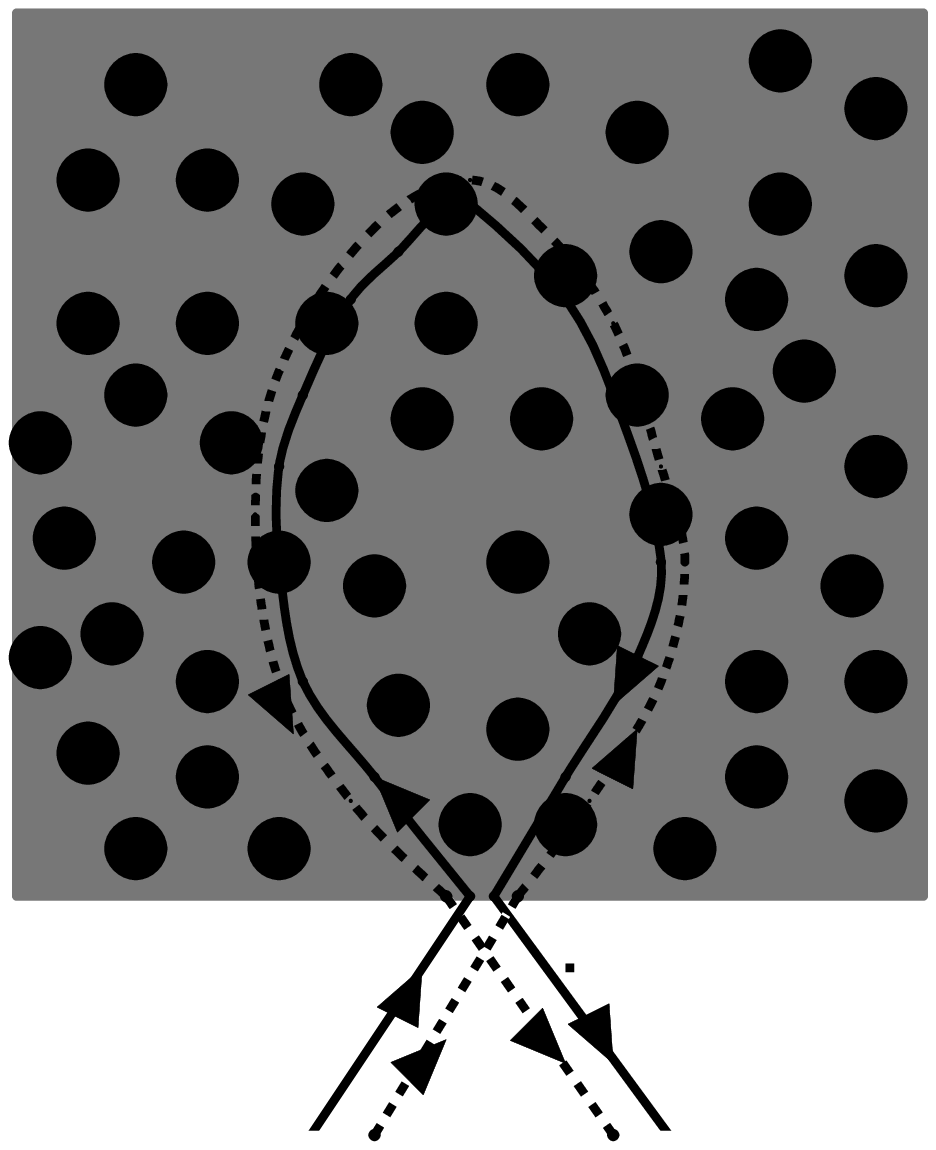}}
\epsfxsize=15cm
\caption{Coherently amplifying(marked by shade) medium in presence of disorder.
Here we emphasize that the coherent-back-scattering (marked by the loop), hence localization
will not be affected due to the presence of coherent amplification.
As a result reflected
wave  will be amplified.  }
\label{ppfig1}
\end{figure}

The main cause of the weak and the strong localization in a
disordered medium is coherent-back-scattering (CBS)
\cite{bergmann,lee}. In the weakly localized regime, CBS is the
correction to the classical transport, and in the strongly
localized regime CBS is the dominant effect. In the presence of
coherent amplification/attenuation, the partial wave amplitudes
get amplified/attenuated in the same way along the echo (i.e.,
time-reversed ) paths when they meet at the starting point 
$r_0$ as shown in Fig.\ref{ppfig1}, i.e., the phase is not lost due to
the coherent nature of the amplification, or attenuation which is
relevant to the stimulated emission of radiation, or the coherent
(stochastic) absorption. Let now briefly describe disordered
coherently amplifying/absorbing media relevant to this problem.
\section{Amplification} 
\subsection{Light localization and coherently amplifying media}
By coherently amplifying media we mean stimulated emission 
of radiation in a lasing medium. Here, we show that the presence of
localization in a highly scattering medium, when the medium is
also an active amplifying one, can \underline{enhance} the
amplification. There is the possibility of using Anderson
localization as a "non-resonant feedback" mechanism for 
lasing action. Recent experiments which show self-sustaining
lasing action in an optically amplifying and highly scattering
medium, without an external cavity, support these aspects which
we will discuss in later section. Although the real situation of
stimulated radiation in a lasing medium involves non-linear
amplification, for the sake of simplicity we will consider a
linear amplifying medium only and discuss enhanced amplification
by the disorder induced confinement. By the linear medium here
we mean the refractive index being independent of the light
intensity. 

\begin{figure}
\epsfxsize=15cm
\centerline{\epsfbox{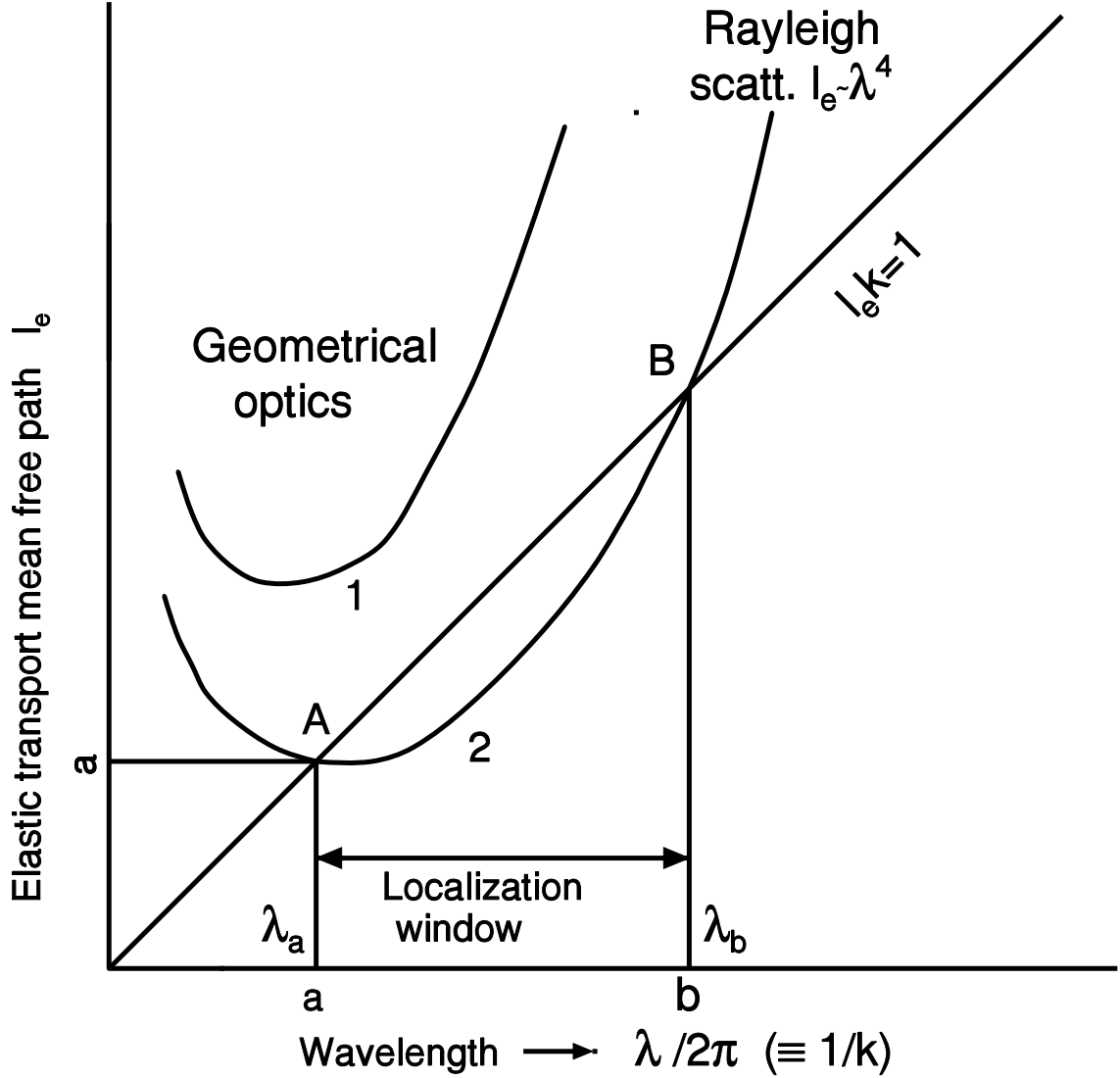}}
\caption{ A schematic picture of the localization window 
for a 3D Maxwell wave in $(k,l_e)$-space, where $k\equiv
2\pi/\lambda$ is the incoming wave vector and $l_e$ is the
free scattering mean free path .
% \protect{\cite{john}}.
Curve 1 is for weak disorder, never satisfy the Mott-Iofee-Regal 
(MIR) criterion, i.e $kl_e\ll 1$.
Curve 2  is for strong disorder case and has a localization window.
 The two mobility edges are
denoted by $a$ and $b$. 
The point $a$ on the $\lambda$ axis is the correlation length of the
disorder. For $\lambda << a $ (geometric optics limit) the MIR
condition for the localization, i.e., $kl_e\ll 1$, is not satisfied. 
For $\lambda >\lambda_b $ (the Rayleigh limit) again the MIR
criterion is not satisfied due to the weak Rayleigh scattering. }
\label{locwindow}
\end{figure}

Light localization is an interesting and clean problem due to the
fact that the energy of the photon is high and 
no effect of temperature and photons are noninteracting.
There are, however, problems in localizing light \cite{ping,john}.
This is because the "effective potential" of a dielectric medium
depends on the incoming wave energy, or the wavelength ($\lambda$).
The problem is to satisfy the Mott-Ioffe-Regel (MIR) \cite{mir}
criterion for the localization which is essential for any wave
localization. The MIR criterion for the localization demands that
$kl_e<<1$, where $k$ is the wave number of the incident wave and
$l_e$ is the elastic (coherent) scattering mean free path for
transport as shown in Fig.\ref{locwindow} . For longer wavelengths,
or in the low energy limit of the incident wave, scattering is
dominated by the Rayleigh scattering, $l_e\sim \lambda^{d+1}$, for a
$d$ dimensional scatterer. In the other extreme limit of high
frequency, or short wave length, the system is almost transparent
(non-scattering) to the incoming wave for wave lengths smaller than
the correlation length of the random potential when the response of
the system is that of geometrical optics. Localization is, thus,
missed in both the wave length limits, the large as well as the small.
While in 1D and 2D all states are localized for arbitrary weak
disorder, there is a very wide localization window where the
localization length is very large. 
In 3D there may be a small free-photon localization window
for intermediate wavelengths. Here the MIR criterion can be
satisfied in an indirect way by first creating a pseudo-band gap
by fulfilling the Bragg condition $\vec k . \vec G = {1\over 2}
\vec G . \vec G $ through a periodic lattice of dielectric
medium and then deliberately introducing a weak disorder to the
system \cite{yprl,johnprl}. It was shown for 3D that in such
systems, the density of states near the band gap correspond to
localized states. 
\subsection{ The questions we are interested }
\begin{figure}
\epsfxsize=17.9cm
\centerline{\epsfbox{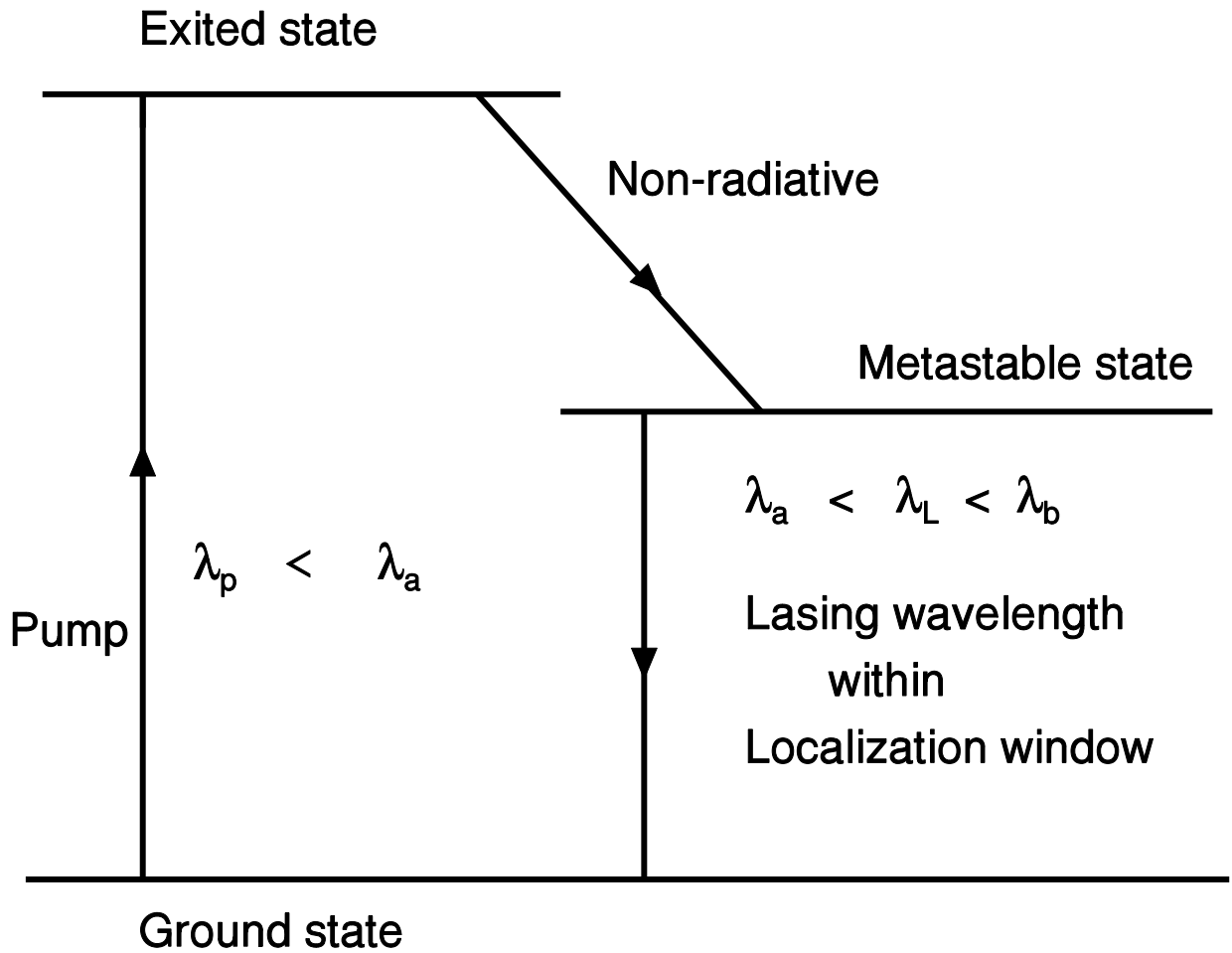}}
\caption{ Proposal for a lasing scheme using  Anderson
localization for a non-resonant feedback mechanism: 
The figure shows a typical 3-level lasing scheme. Proposals for the
lasing action are the following: Keep the pumping wave length
$\lambda_p$, 
which will excite  electrons from the ground state to the excited
state, inside the geometrical optic regime. The wave
length of the lasing signal $\lambda_L$ should lie inside the
localization window, as shown in Fig.\protect{\ref{locwindow}}}
\label{proposal}
\end{figure}

{\bf What will happen to coherent light amplification in an
amplifying (active) medium, when the light modes are already
disorder-localized in that medium?} \\ 
For 1D and 2D all modes are localized for light wave and one can see
the direct affect of localization on coherent
amplification/absorption. In 3D there is a small localization
window and possibility of a physical situation can be as follows.
Suppose the wavelength $\lambda_L$ of a lasing light lies
within the localization window. A typical localization window
for 3D is shown in Fig.\ref{locwindow}. Now, consider a
three-level lasing scheme as shown in  Fig.\ref{proposal}.
For lasing action electrons have to be pumped first from the
ground level to the upper excited levels. The transitions from
the excited level to the intermediate meta-stable level are
arranged to be non-radiative. Lasing action occurs between the
meta-stable and the ground states. Now it is possible to arrange
such that the shorter pumping wavelength $\lambda_p$ that
excites electrons from the ground state to the upper excited
states lies within the delocalized geometric-optical regime,
while the meta-stable to the ground state radiation wave length,
i.e. the lasing wavelength $\lambda_L$, lies within the
localization window. Then, the question is how the localization
and the coherent amplification synergetically lead to enhanced
amplification. Also, how does the lasing wavelength get
selected, if at all in the absence of a resonant cavity
structure providing the feedback.
\subsection{Different feedback mechanisms for lasing action}
In a typical amplifying medium, a wave has to traverse
a long path-length for amplification, by stimulated emission.
Simultaneously, the wave may lose its energy by other loss
mechanisms like non-radiative dissipation, escape out of the
system etc.. The wave requires a positive feedback to gain a
higher amplification. Such a feedback can be given in several
ways. 
\subsubsection{Resonant feedback}
\begin{figure}
\epsfxsize=17.9cm
\centerline{\epsfbox{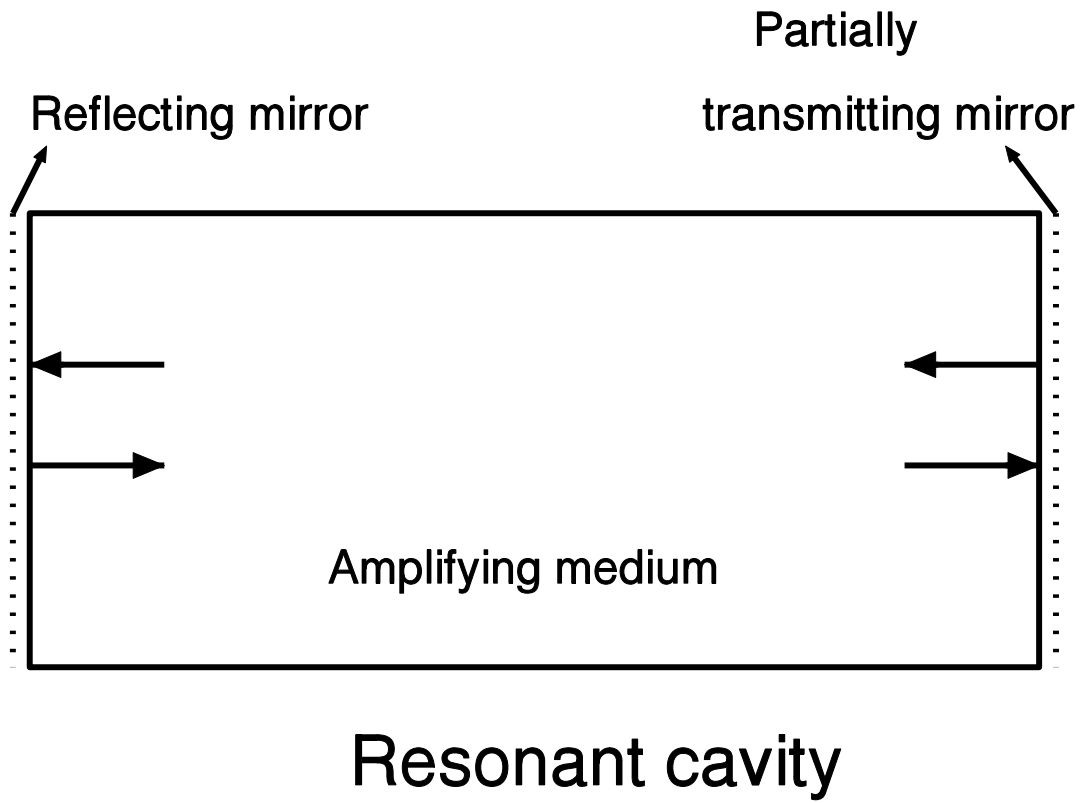}}
\caption{ Schematic picture of a Fabry-Perot etalon, a resonant
cavity for the usual resonant feedback of lasing media. The
amplifying medium is kept between the two parallel mirrors which
are used as resonators. }
\label{fabry}
\end{figure}

In the case of resonant feedback, the laser active
material is kept inside a resonant cavity \cite{laserbk},
which is generally made of two parallel mirrors as shown in
Fig.\ref{fabry}. Stimulated emission has a typical width for its
intensity distribution (gain profile) over the wavelengths.
Resonant cavity is made by matching the spacing between the
two mirrors such that it can give positive feedback selectively
to a set wavelengths that "fit" into the cavity. Given a cavity
of length $L$, we will get the resonant positive feedback,
if $m \lambda/2=L$ is satisfied, where $m$ is an integer. The
wavelength which satisfies the above resonance condition will
survive, because only this wavelength will overcome the energy
losses in the medium through multiple passes. Intensity for the
rest of the wavelengths will die down as there is no feedback to
them to compensate for the loss in the medium. The line-width of
the laser light will now be determined by the cavity ---
Q(quality)-factor. It is much narrower than the width of the
gain profile of the laser material. It is also narrower than the
widths of the cavity resonances. It is in fact "gain-narrowed".
\subsubsection{Non-resonant feedback/feed-forward}
\begin{figure}
\epsfxsize=15cm
\epsfysize=9cm
\centerline{\epsfbox{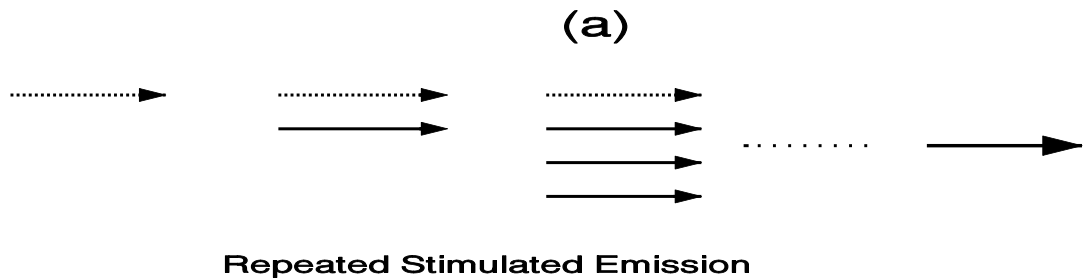}}
\epsfxsize=15cm
\epsfysize=9cm
\centerline{\epsfbox{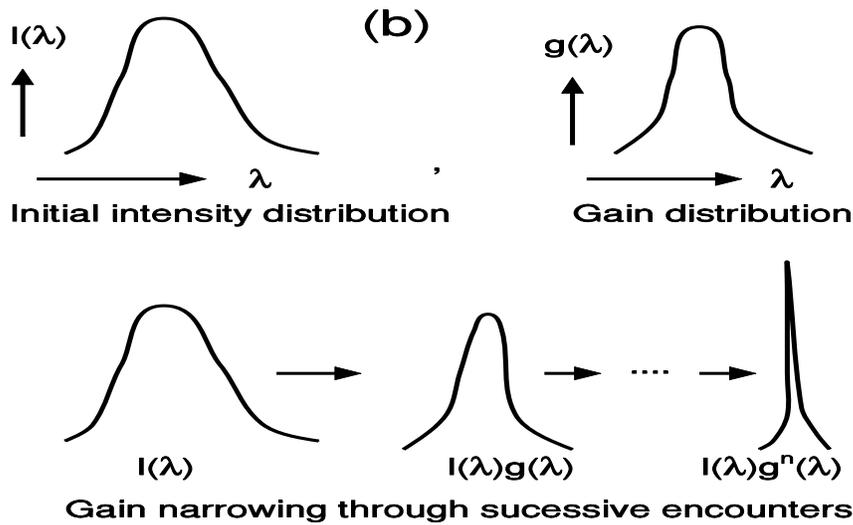}}
\caption{(a) A schematic picture in case of non-resonant "feedback" by
feed-forward in a homogeneous amplifying medium. A wave
is coherently amplified as it moves forward.
The dotted arrows denote the starting wave amplitude
and the solid arrows show the coherent addition to the
wave amplitude by repeated stimulated emission. 
(b) A typical gain narrowing process by successive
coherent amplifications. }
\label{gain}
\end{figure}

A distributed "gain" to a wave propagating in an amplifying medium
can be given to it in its forward journey (single pass) also, if
it moves over a sufficiently long path length as shown in
Fig.\ref{gain}a. An example of this type of medium is to be
found commonly in our galaxy. Astrophysical "maser"
\cite{maserbk}, like the $SiO$ maser in our galaxy gets feedback
by this mechanism. The gain produces its own spectral narrowing
without any resonant structures. In order to see how this happens, 
note that the medium amplification has a gain profile with 
respect to the wavelength. A starting intensity profile, which also
has a distribution over the wavelengths, passing trough the
amplifying medium is modulated by the gain profile. In the
process of a long-length journey, only the prominent peak of the
initial intensity profile will be effectively amplified.
Thus, let $I(\lambda)$ be the starting intensity distribution of
a stimulated radiation and $g(\lambda)$ be the gain profile of
the medium. After $n$ successive coherently amplifying
encounters, the intensity distribution will be amplified
iteratively as:

$I(\lambda)\rightarrow
I(\lambda)g(\lambda)$$\rightarrow$$I(\lambda)g(\lambda)^2$......... 
$I(\lambda)g(\lambda)^{n}$. \\

The latter would tend to a single delta-function as
$n\rightarrow \infty$, even when the initial profile $I(\lambda)$
is broad, but singly peaked.

Fig.\ref{gain}b shows a schematic typical "gain narrowing" by
such a "feed-forward" mechanism. "Feed-forward" is a better
nomenclature here than "feedback".

A variation on the above feed-forward theme is readily realized in a 
diffusing, amplifying medium where random multiple scattering
(turbidity) can enhance the path length traversed by light 
( or increase its residence time) before emerging out of the
amplifying medium. Thus if $l_a$ is gain length for the pure
amplifying medium (i.e., the length over which the intensity is
amplified by a factor $e^{+1}$), and $l_e$ is the transport mean
free path for diffusion due to scattering, then for the
three-dimensional sample size $L \geq L_{SR}\approx \pi
\sqrt{l_gl_e/3}$, the gain exceeds the loss at the boundaries
and self-sustaining emission --- Super Radiance(SR) -- can
result. There is really no feed-back loop here and hence no true
laser action. What one has is the Amplification of uncorrelated
Spontaneous Emission (ASE). It does, however, show gain
narrowing as discussed above. 
\subsubsection{ Non-resonant feedback by Anderson localization :
Mirrorless Random Laser }
We are trying to explore the possibility of using the 
Anderson localized states for a feedback mechanism in the
disordered lasing media. When a wave passes through a disordered
amplifying medium it undergoes coherent multiple scattering. It
is reasonable to think that due to the multiple scattering
processes a wave can get amplified before it escapes from the
system. In a strong disordered media, coherent back scattering
ensures the wave to bring back to its starting point.
 Now the questions are : (a) is this amplification
sufficient to overcome the loss in the system, (b) can we use
the system (amplifying medium with disorder) as a pure optical
amplifier, and (c) is there a mechanism for wavelength
selection. The answers to (a) and (b) seem to be yes, as
supported by recent theories and experiments and answer to (c)
is still unknown. 
%
%\section{Recent experimental observations of lasing action from
%strongly scattering media without an external resonant-cavity}
Recent series of experiments by Lawandy {\sl et. al.}
\cite{lawandy1,lawandy2,prasad}  have
shown evidence for self-sustaining lasing action in
an amplifying, optical scattering medium. Though it seems more
like Amplified Spontaneous Emission (ASE) \cite{lawandy1,wths}.
The experiment was performed on rhodeamine 640 perchlorate laser dye,
dissolved in methanol, as an active medium. Colloidal suspension
of $TiO_2$ nanoparticles (coated with $Al_2O_3$)  in the active
methanol solution was used as the optical scattering random
medium. The emission from such a system was found to exhibit
multi-mode laser oscillations in the absence of any external
resonant cavity. With no $TiO_2$ (no strong disorder) in the
active solution, there was no lasing action from the pure laser
active dye dissolved in methanol. 

Very recent experiment by Wiersma {\sl et. al.} \cite{albada}
shows enhancement of amplification due to coherent back
scattering from an coherently amplifying optical disordered
solid state medium. This experiment is more relevant to our calculation,
which deals with amplification in a medium  due to 
localization. 

These experiments motivate one to study the phenomenon of
amplification-enhancement by disorder in more detail. The
model we are going to consider is for the case of a pure 1D
random amplifying medium. A physical realization may be a single
longitudinal-mode optical fiber doped with a laser active
material ($Er^{3+}$, say) and disordered intentionally. 

\section{ Coherently absorbing media}
There have been recent interest in the role of absorption
on localization, and on the different length scales in the
problem in the presence of absorption. For the bosonic case,
like light etc., a coherent state (e.g. a laser beam) is an
eigenstate of the annihilation operator. Removal of a
photon(absorption) does not destroy the phase coherence.
Fermions cannot be annihilated in the context we are considering
here. For the fermionic case, the physical picturization of the
absorption will be some kind of inelastic process (like
scattering by phonons), where electrons lose its partial temporal
phase memory. This type of absorption is called
{\em stochastic absorption}. Absorption can be of an other type also,
the so called deterministic absorption. In this case,
particles are shut out by a chopper from its beam path. It has
been shown that the stochastic absorption retains more coherence
than the deterministic absorption  while absorbing (removing) 
the same amount of particles \cite{summ}.

Recent theoretical studies  \cite{weaver,yusofin,jayan} have
shown that the absorption in case of light wave does not give
any cut-off length scale for the localization problem. 
If a system is in the localized state, absorption will not kill 
the localization to make the system again diffusive. A sharp
mobility edge exist even in the presence of significant absorption
in 3D.
\section{ Model}
\subsection{Modeling coherent amplification/absorption by complex
potential}
Adding a constant imaginary part with the proper sign to the real
potential of the Maxwell/Schr\"odinger wave  can model the linear
coherent amplification/absorption.
The Maxwell equation can be transformed to the Schr\"{o}dinger 
equation  where the effective potential depends on the frequency of
the incoming wave. Also, both the Maxwell and the Schr\"{o}dinger
equations can be transformed to the Helmholtz equation. 

Adding a constant imaginary potential $iV_a$ to the original real
potential $V(x)$, the Schr\"{o}dinger equation becomes:
\begin{equation}
- \frac{\partial^2 \psi}{\partial x^2} + [ V(x)+iV_a)]  \psi
\, = k^2 \psi \quad,
\label{sceq}
\end{equation}
where  $\hbar^2/2m =1 $. \\
Similarly adding an imaginary constant  term 
$i\epsilon_a$ to the original random refractive index 
$\epsilon_0 +\epsilon(x)$, the Maxwell equation becomes:
\begin{equation}
\frac{\partial^2 E}{\partial x^2} + \frac{\omega ^2}{c^2}
[\epsilon_0 + \epsilon (x) +i\epsilon_a ] E
\, = 0 \quad .
\label{mxeq}
\end{equation}
Here the subscript $'a'$ denotes amplification/absorption(attenuation).

The corresponding Helmholtz equation can be written as:
\begin{equation}
\frac{\partial^2 u}{\partial x^2} +  k^2 [1+(\eta(x)+i\eta_a)] u
\, = 0 \quad,
\label{heq}
\end{equation}
      where  $\eta(x)$=$-V(x)/k^2$,
             $\eta_a$=$-V_a/k^2$ for the Schr\"{o}dinger equation, and
             $\eta(x)$= $+{\epsilon(x)\over \epsilon_0 }$,
             $\eta_a$= $+{\epsilon_a\over \epsilon_0}$ 
and $k^2={\omega^2\over c^2}\epsilon_0$ for the Maxwell equation.
Here, $\epsilon_0 $ is the constant dielectric background and
$\epsilon(x)$ is the randomly spatially fluctuating part of the
dielectric constant. $k\equiv 2\pi/\lambda$, where $\lambda =$
wavelength in the average medium $(\epsilon_0)$.

The model we are considering here is for one-channel scalar wave
where the polarization aspects have been ignored. This holds good for a
single-mode polarization maintaining optical fiber.
Formally, we will consider 1D one-channel Helmholtz equation
only, and will specify at the proper place which situation is
appropriate, optical or electronic. 

It has been pointed out by Rubio and Kumar \cite{rubio}
that modeling absorption by complex potential will always 
give a concomitant reflection. Increasing the strength of the
complex potential will not increase the absorption. There is
a competition between the amplification/absorption and the
reflection. For very high absorption, the system may try to act
as a perfect reflector. Later we present a detailed analysis for
the modeling of absorption by "fake", or "side", channels
obviating the need for a complex potential. 
\subsection{ The Langevin Equation for the complex amplitude
reflection coefficient R(L)}
The Langevin equation for the complex amplitude reflection
coefficient $R(L)$ can be derived by the invariant imbedding
method \cite{rammal} from the Helmholtz equation Eq.\ref{heq},
\begin{equation}
  \frac{d R(L)}{d L} \, = \,\,  
2ik R(L)\, + \,i \frac{k}{2} (1+\eta (L) + i \eta_a )(1+R(L))^2 \quad,
\label{leq}
\end{equation}
 with the initial condition R(L)=0 for L=0.
\subsection{The Fokker-Planck Equation }
Now, taking $R(L)= \sqrt{r} e^{i\theta }$, the Langevin
equation Eq.\ref{leq} reduces to two coupled differential
equations. 
\begin{eqnarray}
\frac{d r}{d L}
\, &=&   k \eta (L)  r^{\frac{1}{2}}(1-r) \sin\theta 
      - 2k\eta_a r \nonumber \\
&-&  k \eta_a r^{\frac{1}{2}}(1+r)\cos\theta   \\
\frac{d \theta}{d L} &=&
2 k + {k\over 2} \eta(L)\left[ 2 + cos\theta ( r^{1/2} + r^{-1/2})\right]
  \nonumber \\
 &-& {k\over 2} \eta_a ( r^{1/2} - r^{-1/2} )\sin\theta \quad.
\end{eqnarray}
The same way as described in Ref.\cite{rammal}, using stochastic
Liouville equation for the evolution of probability density and then
integrating out the the stochastic part by Novikov's theorem, we
get the Fokker-Planck equation: 
\begin{eqnarray}
\frac{ \partial P(r,\theta) }{\partial l} &=& 
  \left[ \sin\theta\frac{\partial}{\partial r}( r^{1/2} (1-r))
 + \frac{\partial }{\partial \theta}  \right. \nonumber \\
&& \left.+ {1\over 2} (r^{1/2} + r^{-1/2}) \frac{\partial}{\partial \theta} 
    cos(\theta) \right ]^2 P(r,\theta)  \nonumber \\
&& - 2 k\xi \frac{\partial}{\partial \theta} 
\left[ P(r,\theta ) \right]  \nonumber \\
&& + D  \frac{\partial}{\partial r} 
\left[ r P(r,\theta ) \right]  \nonumber \\
&& + {D\over 2} \cos \theta \frac{\partial }{\partial r} 
\left[ r^{1/2}(1+r) P(r,\theta ) \right]  \nonumber \\
&& + {D\over 4}  (r^{1/2}-r^{-1/2})
\frac{\partial}{\partial \theta } \left[\sin \theta P(r,\theta ) \right]
\quad, 
\label{fpeq}
\end{eqnarray}
where we have introduced the dimensionless length
$l= \frac{L}{\xi}$  and 
$D=\frac{4\eta_a}{qk}$ = $\xi/\xi_a$, and $\xi\equiv ({1\over
2}qk^2)^{-1}$ is the localization length and $\xi_a={1\over
2k\eta_a}$ is the active (amplifying/absorbing) length scale in
the problem. 
\subsection{Parameter scales in the problem}
The above Fokker-Planck equation (Eq\ref{fpeq}) has effectively three
dimensionless scales:\\ 
(1) Length $l=L/\xi$, where $L$ is the length of the sample and
  $\xi=({1\over 2} q k^2)^{-1}$ is the localization length,
 $k$ is the incoming wave vector and $q$ is the delta-correlation 
strength of the potential.\\
(2) Disorder parameter $2k\xi$.\\
(3) Active parameter $D=\xi/\xi_a$, where $\xi_a={1\over (2 k\eta_a)}$ is
the active length scale. \\
In the case of amplification, $\xi_a$ is usually known as the
gain length $l_g$. The gain length $l_g$ decreases with
increasing population inversion(pumping), and eventually
saturates at high enough pumping rate. 
\section{  Results and Discussion }
\subsection{ Analytical solution of the Fokker-Planck equation 
within the random phase approximation (RPA) } 
The Fokker-Planck equation Eq.\ref{fpeq} can be solved analytically
for the asymptotic limit of large lengths in the random phase
approximation (i.e., for the weak disorder case). Integrating
the phase part of the FP Eq.\ref{fpeq}, i.e., $P(r)= 1/(2\pi)
\int P(r,\theta) d\theta $, one gets FP equation of the
reflection coefficient $r$: 
\begin{eqnarray}
\frac{\partial P(r,l)}{\partial l} \, &=& \, r(1-r)^2 
\frac{\partial^2 P(r,l)}{\partial r^2} \, \nonumber \\
&+& \, [1+(-6+D)r + 5r^2]
\frac{\partial P(r,l)}{\partial r} \nonumber \\
&+& [(-2+D) + 4r]P(r,l)\quad ,
\label{fpeqr}
\end{eqnarray} 
(To make the point more clear we have written $P(r)$ as $P(r,l)$
to show the explicit length-dependence)

Here, $D < 0$ corresponds to coherent amplification,
$D > 0$ corresponds to coherent absorption, and $D = 0$ 
corresponds to the unitary case. To emphasize the dependence of
$P(r,l)$ on $D$ we will write here $P(r,l) \equiv
P^D(r,l)$ . 

Few asymptotic features of $P^D(r,l)$ can be obtained
directly from Eq.(\ref{fpeqr}). We can note that Eq.
(\ref{fpeqr}) reduces to the unitary case, for $(D=0)$ in the
limits $\eta_a=0$ trivially, and large $q$ nontrivially.
The latter case is dominated by strong disorder and the localization
length is too short for the wave to have penetrated the
amplifying medium. Then, the statistics $P^D(r,l)$
have steady state distribution for $l\rightarrow \infty$. It
can be obtained by setting $ \partial P(r,l)/\partial l = 0 $
and solving the resulting equation analytically. We get the
following steady state distributions:
\newpage
\begin{mathletters}
\begin{eqnarray}
&& \nonumber \\
{\em %\bullet
 (a)\,\,Amplifying\,\, media:} && \nonumber \\
\lim_{l\rightarrow\infty\atop D < 0}
 P^D(r,l) & = & \left\{\begin{array}{cl}
{|D| exp(-{|D|\over r-1})\over(1-r)^2} & \quad \mbox{for}\; r\geq 1\\
0 &\quad\mbox{for}\; r < 1,\end{array}\right. 
\label{EQ4A}\\
%\eeq
&& \nonumber \\
{\em %\bullet
(b) \,\, Absorbing\,\, media:} && \nonumber \\
%\beq
\lim_{l\rightarrow\infty\atop D > 0}
  P^D(r,l) & = & \left\{\begin{array}{cl}
\frac{|D| exp(|D|) exp(-{|D|\over 1-r})}{(1-r)^2} &
 \quad \mbox{for}\; r\leq 1\\
0 &\quad\mbox{for}\; r > 1,\end{array}\right.\nonumber \\
\label{EQ4B}\\
%\eeq
 && \nonumber \\
{\em %\bullet 
(c)\,\, Passive\,\, media:} && \nonumber \\
%\beq
\lim_{l\rightarrow\infty\atop D = 0}
  P^D(r,l) & = & \delta(1-r)
\label{EQ4C}
\end{eqnarray}
\end{mathletters}

\begin{figure}
\epsfxsize=17.9cm
\centerline{\epsfbox{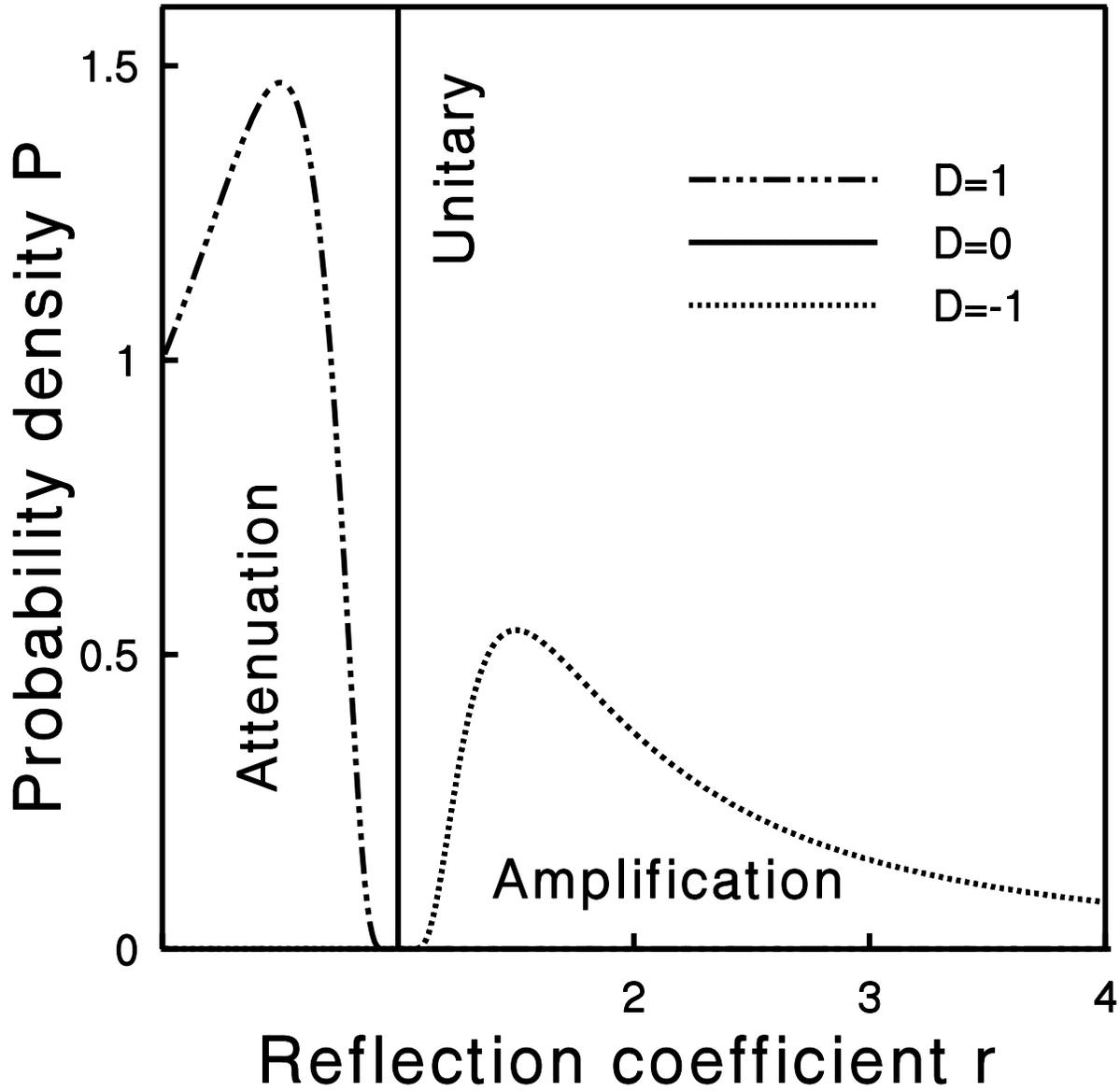}}
\caption{ Limiting probability distribution $P^D_{l\rightarrow\infty}$
of the reflection coefficient $r$, within the random phase
approximation, for the cases 
(a) Coherent absorption ($D=1$),
(b) Unitary ($D=0$), and
(c) Coherent  amplification ($D=-1$).}
\label{plimit}
\end{figure}

In Fig.\ref{plimit} we have plotted $P^D(r,l=\infty)$ for the
three limiting cases with parameter $ D= +1,\, 0,\, -1$. 
{\em These steady state probability distributions for $P(r)$ with the
weak disorder show that for the case of amplifying media,
the probability distribution of the reflection coefficient
effectively lies within $r>1$ region and spread out to infinity. In
this asymptotic limit of large sample length the average $<r>$
diverges for any $D$, i.e, we have superradiance in fact
laser action.}

Fig.\ref{plimit} clearly shows that the Anderson localization 
enhances the amplification. Thus, there is a possibility that
the Anderson localization can be used for the non-resonant
feedback mechanism for lasing action.  These probability 
distributions have universal nature and have been studied
recently in more detail using different approach
by several  authors 
\cite{zq1,zq2,zq3,zq4,aj1,aj2,jh,cwj1,cwj2,zy,fr1,fr2,aks,kim}.
\subsection{ Numerical solution of the Fokker-Planck equation
beyond the random-phase-approximation:  Marginal probability
distribution $P(\theta)$ of the phase $\theta$ and the probability
distribution $P(r)$ of the reflection coefficient $r$.} 
The Fokker-Planck equation, Eq.\ref{fpeq}, is difficult to
solve analytically beyond the random phase approximation.
We  solve the full Fokker-Planck equation (\ref{fpeq}) numerically
to investigate the effects of phase-correlation on the
statistics of the reflection coefficient. Initial probability
distribution $P(r,\theta)$ at $l=0$ for our numerical work was
taken to be the same as in the case of the passive disordered
media described in our previous paper \cite{pp}. 
Now, $r$ in case of absorption is limited between 0 and 1, and in
case of amplification, $r$  is limited between 0 and $\infty$. 
For our numerical calculation for the amplifying case,
however, we have considered values of $r$ sufficiently large but
finite  such that the truncated range of the probability
distribution shows the main feature of the full distribution.
Below we are giving in detail, the simulation results of the
Fokker-Planck  equation in all important parameter regimes.
\subsubsection{ Phase distribution with active parameter D,
for the case of weak disorder } 
\begin{figure}
\epsfxsize=17.9cm
\centerline{\epsfbox{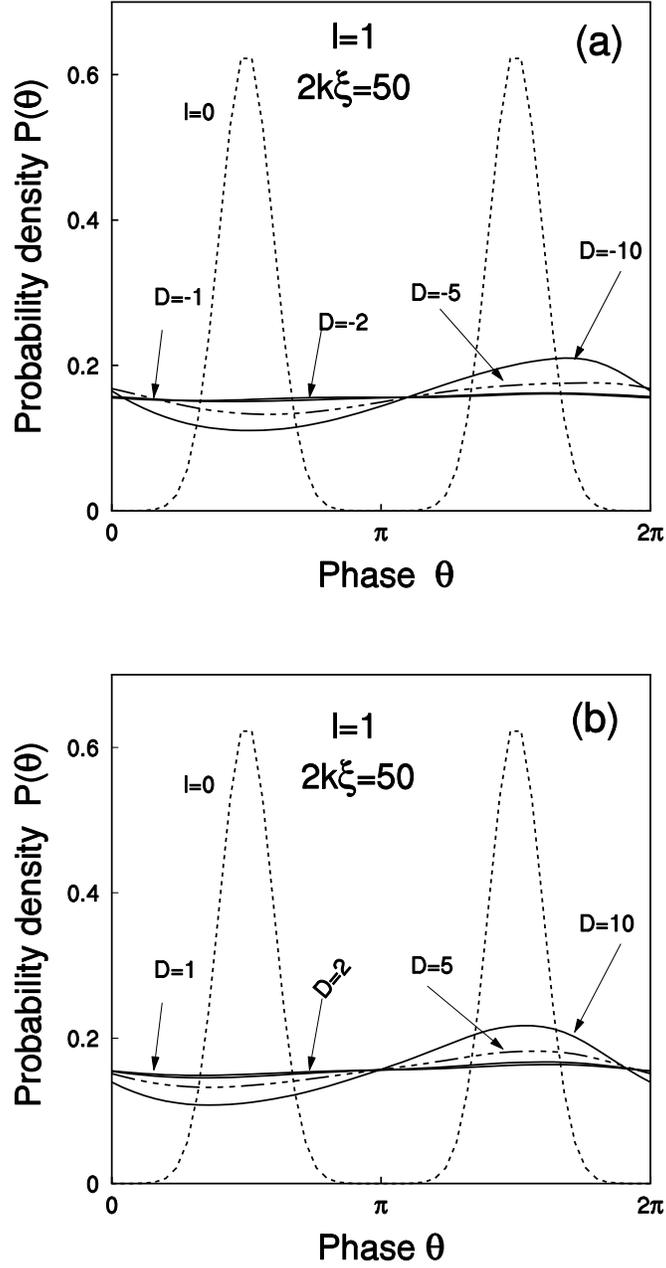}}
\caption{ Probability distribution $P(\theta)$ against the
active parameter strength $D$ in the weak disorder regime 
for a  fixed disorder parameter strength $2k_F\xi=50$ and 
fixed sample length $l=1$. Plots are for
(a) Coherent amplification (D are -ve), and 
(b) Coherent absorption (D are +ve).}
\label{fig8}
\end{figure}

 In Figs.\ref{fig8}(a),(b) we have plotted 
 the variation of the phase in the
random phase approximation with the active parameter $|D|$, with
fixed disorder parameter $2k\xi=50$ (weak disorder regime) and
sample length $l$. It is clear that the random phase
approximation holds up to a large value of $|D|$, for both (a)
amplifying and (b) absorbing medium.  At this  point we can
tell that the demand of the Ref.\cite{aj1,aj2} on the issue of phase
distribution is invalid. Once disorder
is weak, the RPA persist to a higher amplification/absorption
values.
\subsubsection{Weak disorder with weak amplification/absorption}
\begin{figure}
\epsfxsize=17.9cm
\centerline{\epsfbox{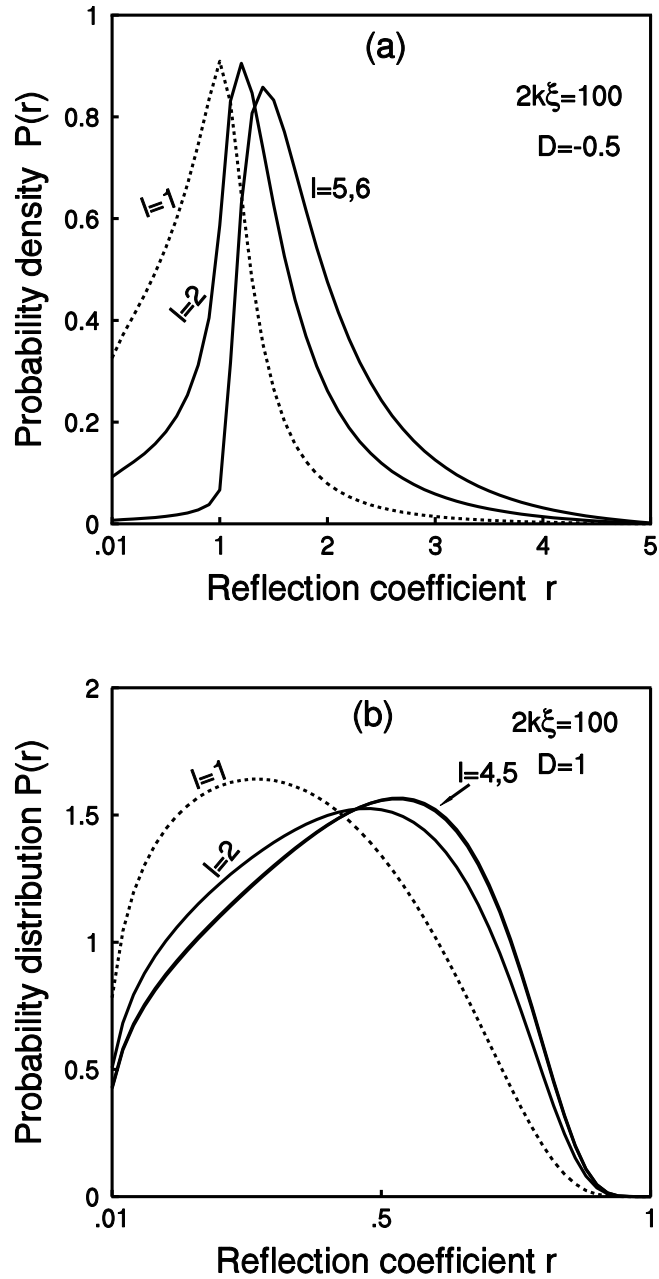}}
\caption{ Probability distribution $P(r)$ against the sample
length $l$, with fixed  \underline{weak} disorder strength $2k_F\xi
=100$ and fixed \underline{weak} active parameter $|D|$.
Plots are for 
(a) Coherent amplification (D = -0.5), and
(b) Coherent absorption   (D = +1).}
\label{fig9}
\end{figure}

Figs.\ref{fig9} (a),(b) are the plots of the reflection coefficient $r$
distribution $P(r)$ for weak $|D|$. These plots show the evolution
with the length of the sample and saturation for large lengths.
The distribution matches the analytical result for the
asymptotic limit of large sample lengths within the random phase
approximation for both (a) the amplifying and (b) the absorbing
cases. 
\subsubsection{Weak disorder with strong amplification/absorption}
\begin{figure}
\epsfxsize=17.9cm
\centerline{\epsfbox{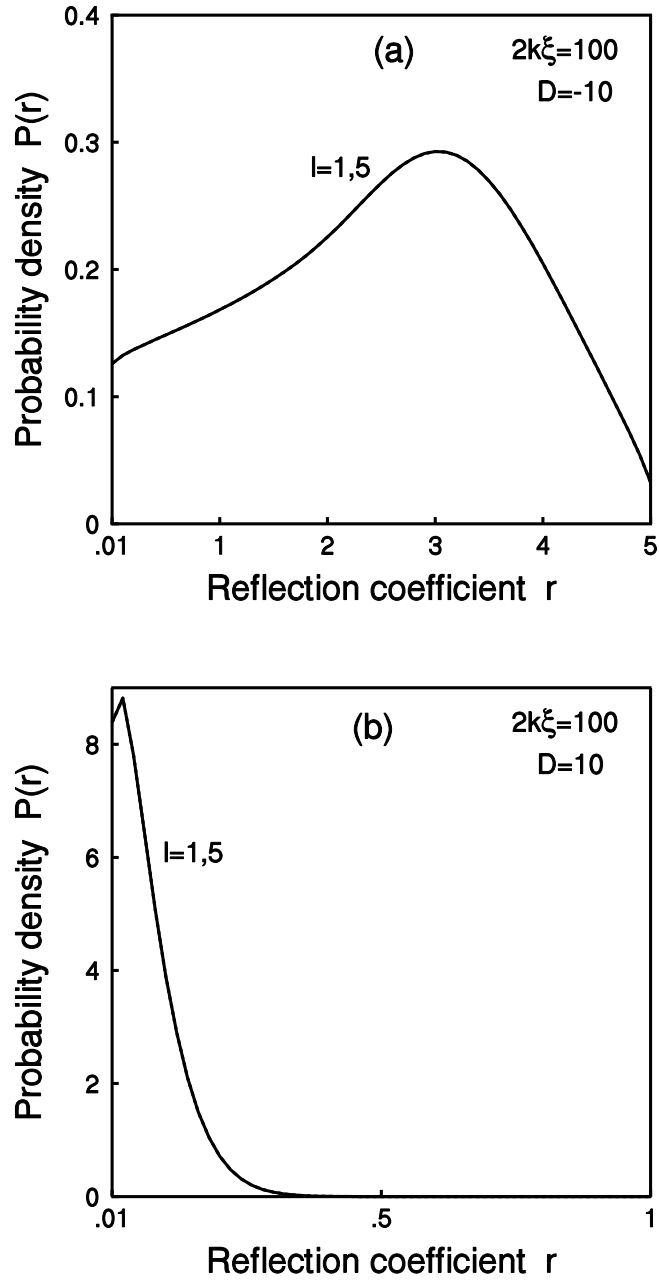}}
\caption{Probability distribution $P(r)$ against the sample length $l$,
with fixed  \underline{weak} disorder strength $2k_F\xi
=100$ and fixed \underline{strong} active parameter $|D| =10$.
Plots are for the cases 
(a) Coherent amplification (D = -10), and
(b) Coherent absorption   (D =  +10). }
\label{fig10}
\end{figure}

Figs.\ref{fig10}(a),(b), are the plots of $P(r)$ for different
lengths. (a) In the case of an amplifying medium $P(r)$ has 
complicated behavior. (b) In the case of absorbing media,
the waves get absorbed before it gets reflected  which explain
the peak of the $P(r)$  near $r=0$. 
\subsubsection{ Strong disorder with weak amplification/absorption }
\begin{figure}
\epsfxsize=17.9cm
\epsfysize=16cm
\centerline{\epsfbox{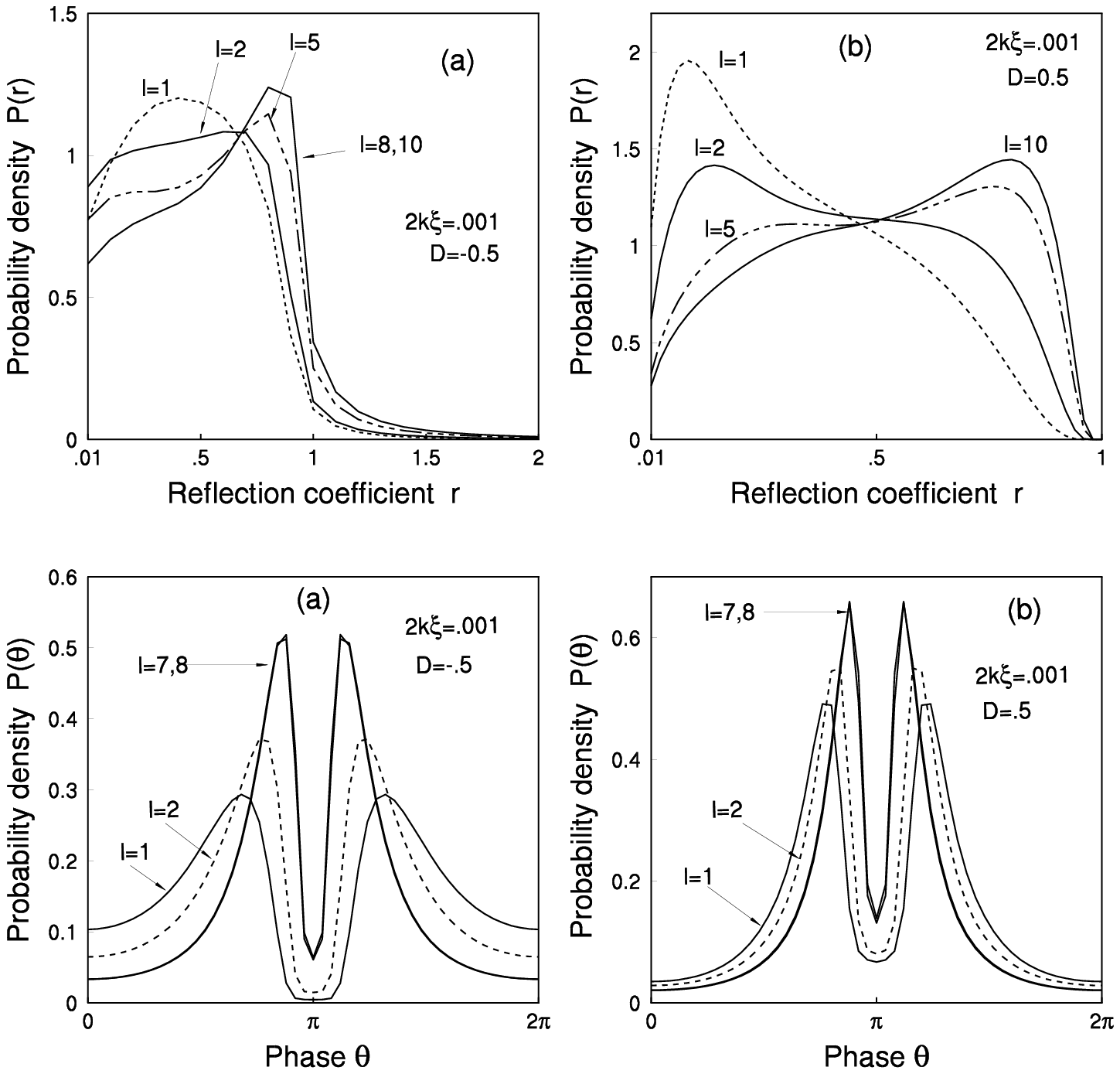}}
\caption{ Probability distribution $P(r)$ and $P(\theta)$ separately
against the sample length  $l$. 
Parameters are fixed \underline{strong} disorder strength $2k_F\xi
=.001$ and fixed \underline{weak} active parameter $|D|= 0.5$.
Plots are for the cases
(a) Coherent amplification ($D = -0.5$), and
(b) Coherent absorption ($D = +0.5$). } 
\label{fig11}
\end{figure}

In Figs.\ref{fig11}(a),(b), two top plots are the evolution of $P(r)$
with different sample lengths $l$ for (a) Amplification and (b)
Absorption. Competition between the 
localization and amplification/absorption is implied by the double
peaked distribution. The signature of more localization is that
the reflection probability will  peak near $r=1$ and the
signature of more absorption is that P(r) will  peak near
$r=0$. 

 In Figs.\ref{fig11}(a),(b), bottom two plots are the phase
distributions $P(\theta)$ correspond to the $P(r)$ distribution
for (a) amplifying and (b) absorbing cases. The $P(\theta)$
distributions are essentially like those of
 the passive medium \cite{pp} ,
peak near $\theta=\pi$ symmetrically and do not get distorted
much. Cases are for: (a) Amplification and (b) Attenuation. 
\subsubsection{Strong disorder with strong amplification/absorption}
\begin{figure}
\epsfxsize=17.9cm
\epsfysize=16cm
\centerline{\epsfbox{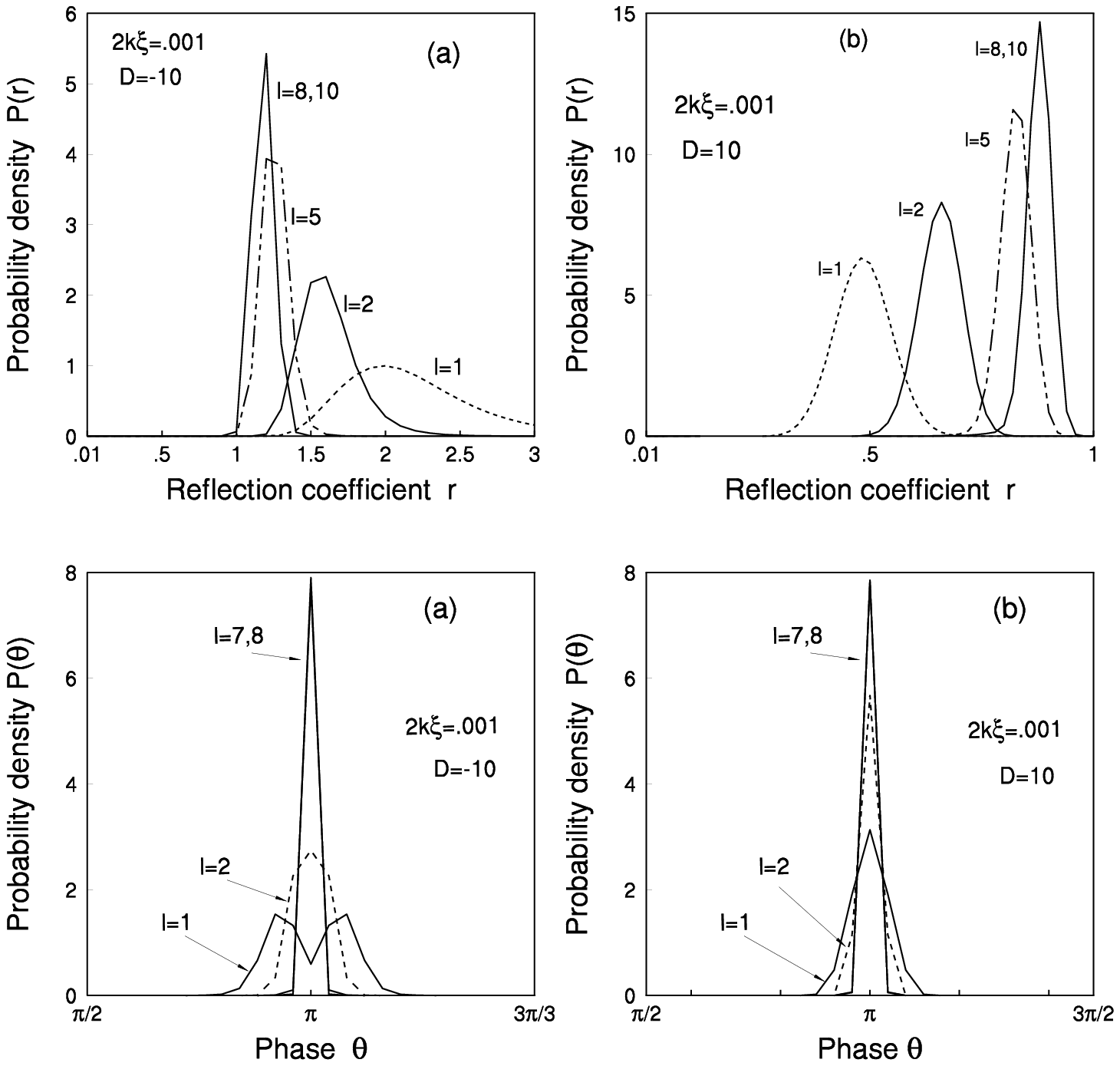}}
\caption{ Probability distribution $P(r)$ and $P(\theta)$ separately
against the sample length  $l$. 
Parameters are fixed \underline{strong} disorder strength $2k_F\xi
=.001$ and fixed \underline{strong} active parameter $|D|= 10$.
Plots are for the cases 
(a) Coherent amplification ($D = -10$), and 
(b) Coherent absorption ($D = +10$) }
\label{fig12}
\end{figure}

In Figs.\ref{fig12}(a),(b), top two plots have the following features:\\
(a) for the amplifying case the probability spreads to the right
side of $r=1$ for $l=1$ and peak of the probability $P(r)$ slowly moves
towards $r=1$ for larger lengths.\\ 
(b)  shows that for the absorbing media, the probability
initially peaks near $r=0$ and the peak slowly moves towards
$r=1$ for larger lengths.

In Figs.\ref{fig12}(a),(b), bottom two plots show that the steady state
phase distributions is a delta function peaking symmetrically about
$\theta=\pi$, for both the cases (a) amplifying and (b) absorbing.
 This result contradict the recent results of Ref.\cite{aj1,aj2}
   
These results (Figs. \ref{fig12}(a),(b)) support
that the system is behaving as a perfect reflector. It was
indicated by Rubio and Kumar \cite{rubio}, that modeling with a
complex potential is always associated with  reflection due to 
the mismatch of potentials (real and complex). Increasing the
strength of the imaginary part of the potential will not
necessarily increase the amount of absorption or amplification,
i.e., absorption/amplification is not a monotonic function of the
strength of the  {\em complex potential}. Here we show by
explicit numerical simulations that the system 
 behaving as a perfect
reflector for higher lengths, with strong disorder and strong
active parameter $|D|$.  We will discuss this issue next.\\
%We  derive an expression for the Langevin equation taking only 
%the absorption part into consideration and without considering
%the reflection part associate with the absorption, for the case of
%an absorbing media. \\ 
%\vsapce{2cm}

\section{ Stochastic Absorption}
\subsection{ Stochastic Absorption from an Absorbing Side
("Fake") Channel (modeling absorption without reflection)} 
We have shown and discussed that the absorption is not possible
without reflection in a model calculation where 
amplification/absorption is modeled by adding a constant
imaginary potential to the real potential. In the Langevin
equation for $R(L)$, derived from these types of model, the wave
always gets a reflected part along with the absorption. 
There is, however, another way of deriving a Langevin equation for
the reflection amplitude $R(L)$ such that the absorption does
not have a concomitant reflected part. This approach is
motivated by the work of Buttiker \cite{buttiker,maschke} where some
purely absorptive "fake" (side) channels are added to the 
purely elastic scattering channels of interest. A particle that
once enters to the absorbing channel, never comes back and it is
physically lost. In the Appendix, we have
derived the Langevin equation for $R(L)$ following the
B\"{u}ttiker approach. This Langevin equation has some formal
differences from the Eq.\ref{leq}. 
\subsection{ The Langevin Equation }
The Langevin equation for the stochastic absorption only
turns out to be  :
\begin{equation}
  \frac{d R(L)}{d L} \, = \, - \alpha R(L) +  
2ik R(L)\, + \,i \frac{k}{2} (\eta (L))(1+R(L))^2 \quad ,
\label{leqst}
\end{equation}
(for derivation and discussion see Appendix)
with the initial condition R(L)=0 for L=0, and $\alpha$ is the
absorbing parameter.
\subsection{The Fokker-Planck Equation}
Similarly, following the same steps as we discussed previously,
from the Langevin Eq(\ref{leqst}) we derive the Fokker-Planck
equation: 
\begin{eqnarray}
\frac{\partial P(r,\theta)}{\partial l} &=&
 \left[ \sin \theta \frac{d}{dr}( r^{1/2} (1-r))
       +   \frac{\partial}{\partial \theta} \right. \nonumber \\
 &+& \left. {1\over 2} (r^{1/2} + r^{1/2}) \frac{\partial}{\partial \theta} 
     (  cos(\theta))\right] ^2 P(r,\theta) \nonumber \\
 &-&2 k\xi \frac{\partial P(r,\theta)}{\partial \theta}
 + D  \frac{\partial(Pr)}{\partial r} \quad ,
\label{fpst}
\end{eqnarray}
where parameters $l$ and $2k\xi$ are same as defined in Eq.\ref{fpeq},
and $D={4\alpha\over qk}$.
\subsection{ Solution of the FP (Stochastic absorption case)
equation within the Random Phase Approximation } 
Eq \ref{fpst} gives the same Fokker-Planck equation in $r$,
i.e., Eq. \ref{fpeqr} for the reflection coefficient when phase
part is integrated out by the RPA in the limit of weak disorder.
Hence, steady state solutions are also same as Eq.\ref{fpst}.
\subsection{ Numerical solution of the FP equation for the case of
stochastic absorption beyond the RPA for strong disorder}
For the weak disorder case, where random phase approximation
is valid, results are the same as for modeling with a complex
potential. 
We will consider here only the strong disorder case for
the numerical calculations of the Fokker-Planck equation
\ref{fpeq} within the RPA. 
\subsubsection{Strong disorder and Weak Absorption}  
\begin{figure}
\epsfxsize=17.9cm
\centerline{\epsfbox{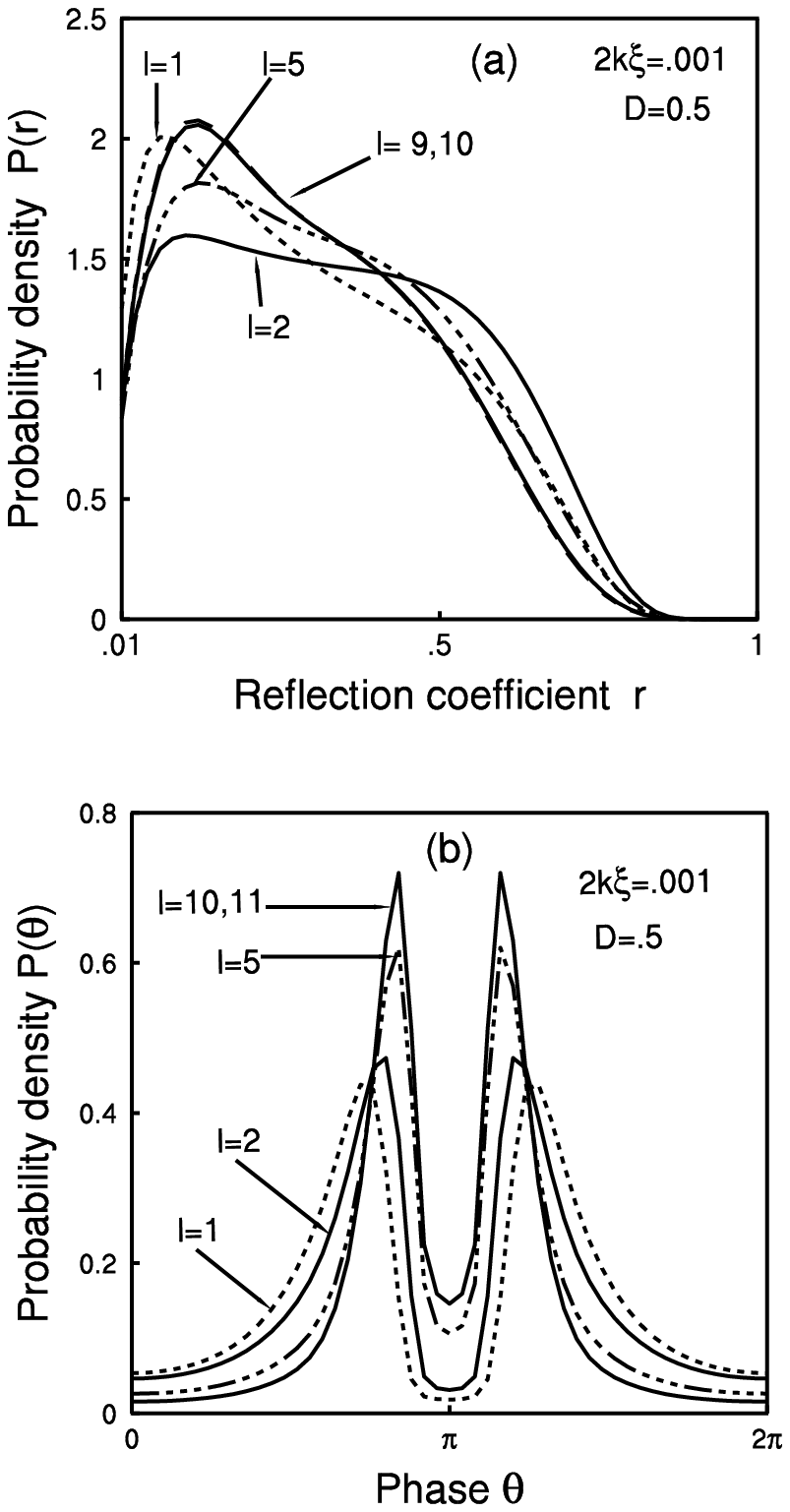}}
\caption{ Probability distributions (a) $P(r)$ and (b) $P(\theta)$
against the sample length $l$ with
fixed \underline{strong} disorder strength $2k_F\xi =.001$ and
fixed \underline{weak} stochastic absorption parameter $D= 0.5$.}
\label{fig13}
\end{figure}

Fig.\ref{fig13}(a) is the plots of $P(r)$ and $P(\theta)$ ,
respectively, for the strongly disordered media with weak stochastic
absorption. Probability distribution for the reflection
coefficient has double peaked behaveour which arises due to
the competition between the absorption and localization.

Fig.\ref{fig13}(b) is the plot of $P(\theta)$ which is typical of
a double peaked symmetric distribution, similar as that for
strongly absorbing and strong disordered media. 
\subsubsection{Strong disorder and Strong Absorption}
\begin{figure}
\epsfxsize=17.9cm
\centerline{\epsfbox{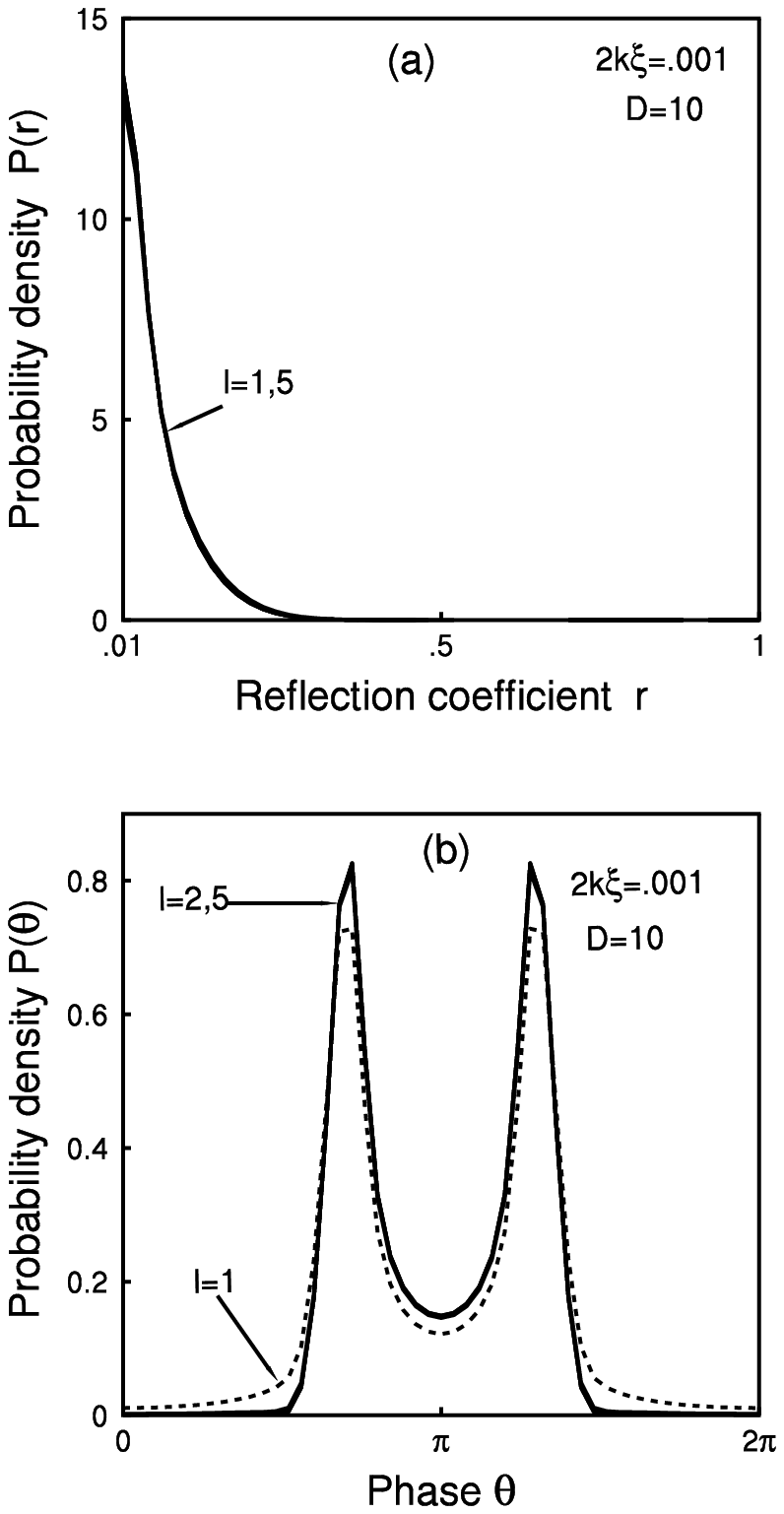}}
\caption{ Probability distributions (a) $P(r)$ and (b) $P(\theta)$
against the sample length $l$ with
fixed \underline{strong} disorder strength $2k_F\xi =.001$ and
fixed \underline{strong} stochastic absorption parameter $D= 10$.}
\label{fig14}
\end{figure}

Fig.\ref{fig14}(a) is the plot of $P(r)$. The wave is
absorbed in the medium before it gets reflected as we have
modeled absorption without reflections and we are considering
the case of strong absorption, and hence the probability of
absorption is more than that of reflection. 

Fig.\ref{fig14}(b) shows the phase distribution $P(\theta)$,
which is a double peaked symmetric distribution and differs from
the model of absorption by imaginary potential, which was a
delta function distribution at $\theta=\pi$.
%
%\section{ Recent theoretical works on random amplifying
%media } 
\section{ Discussion and Conclusions}
There are a few recent theoretical works on random amplifying
media for the weak and the strong disordered regimes.
Zhang \cite{zq1,zq2,zq3} has simulated the 1D disordered
amplifying/absorbing medium by transfer matrix method and got
the same steady state distributions for $P(r)$ as ours,
thus confirming our analytical treatment.
Recently,  calculation of Beenakker {\sl
et. al.} \cite{cwj1,cwj2} have solved the 1D, N-channel 
case with amplification/absorption, both analytically and 
numerically, using the Dorokhov-Mello-Pereyra-Kumar
(DMPK) \cite{dmpk} equation. Their study shows that for the
N-channel case with absorption, reflection probability
distributions are Gaussian; and for the amplifying case
reflection probability follow the Laguerre distribution. For
$N=1$, these equations reduce to ours.
%Zyuzin \cite{zyuzin} has calculated the amplification in a
%disordered sample in the weak localization regime. His
%calculation shows the enhancement of the amplification due to
%the coherent back scattering. 
All the theoretical studies 
\cite{zq1,zq2,zq3,zq4,aj1,aj2,jh,cwj1,cwj2,zy,fr1,fr2,aks,kim}
conclude that the amplification
is enhanced by the localization effect of  disorder.

In conclusions, we have calculated the statistics of the reflection
coefficient $(r)$ and its associated phase $(\theta)$ for a wave
reflected from an amplifying/absorbing disordered medium, for
different disorder strengths and lengths of the sample.
This numerical calculation clarifies several doubts recently
raised on our previous paper \cite{ppnk} on random phase approximation.
We have
modeled coherent amplification/absorption by adding a constant
imaginary part to the real potential. 
We have discussed the cases of modeling  with a complex potential.
Following the Buttiker S-matrix approach, we have derived a
Langevin equation for the reflection coefficient, which models
absorption without the reflection and represent a case of pure
stochastic absorption. Within the RPA, the FP equation for the
case of stochastic absorption is the same as that for absorption
modeled by a complex potential.  Drawback of this numerical model
calculation is that we can not go very large $"r"$ due to the computation
limitation for amplifying case, and for this reason we have not calculated
the average $"r"$

Our conclusions are the following:\\
{\em Anderson localization enhances the amplification 
/attenuation in a coherently amplifying/absorbing disordered media.
The effective length of the pseudo cavity of the lasing action
is the order of the localization length.}

{\em Modeling active medium with a complex potential gives the
following details results for both the amplifying and the
absorbing cases:} 

(a) ${\bf Weak}$ disorder and ${\bf Weak}$ active strength:\\
Random-phase approximation is good for weak disorder ($2k\xi >>1$).
The strength of the active part does not affect the random phase
approximation up to a high value of the active parameter (D).
Probability distribution $P(r)$ of the reflection coefficient
$r$ has steady state solutions.

(b) ${\bf Weak}$ disorder and ${\bf Strong}$ active strength:\\
The random phase approximation (RPA) still holds for this region
($2k_F\xi >>1$). Strength of the active part does not significantly
affect the random phase approximation. $P(r)$ saturates
very fast. Wave gets absorbed before the reflection.

(c) ${\bf Strong}$ disorder and ${\bf Weak}$ active strength:\\
Random phase approximation does not hold in this regime.
The strength of the active parameter does not affect much the phase
distribution, which remains similar to that for the passive medium. 
There is a competition between amplification/absorption
and localization which is shown by the double peaked
distributions of $P(r)$.

(d) ${\bf Strong}$ disorder and ${\bf Strong}$ active strength:\\ 
The probability distribution of the reflection coefficient 
 peaks  near r=1, for large lengths.
The steady state reflection distribution $(P(r))$ is
basically a delta function at $r=1$ and the steady state phase
distribution $P(\theta)$ is a delta function at $\theta = \pi $,
i.e., total reflection with opposite phase. The medium behaves as
a perfect reflector. 
        
(e) {\bf Stochastic absorption:}\\
Modeling  {\em absorption without reflection} shows that for the weak
disorder case, the model is same as that of modeling absorption by
adding an imaginary part to the real potential. 

(i) In case of {\bf strong} disorder, and {\bf weak}
absorption, phase distribution $P(\theta)$ does not change
relative to the passive disordered medium and the probability
distribution of the reflection coefficient is similar to the
model with the complex potential. The phase distribution
typically has double peaks similar to that for a passive medium     
 \cite{pp}.

(ii) For {\bf strong} disorder and {\bf strong} absorption, the wave
cannot penetrate deep inside the sample; it gets absorbed before it 
is reflected back. $P(r)$ peaks near $r=0$ and the phase-distribution
tries to peak far from $\theta=\pi$, which is different from
modeling absorption by introducing a complex potential.

Finally, it would be interesting to study the
problem in higher dimension with nonlinear amplification, we are
looking for this. This problem also demand more experiments to look
for. As fine optical fibers are easily available, it may not be
difficult to do the experiment at lower dimensions.  
\appendix
\section{Stochastic Absorption }
\begin{figure}
\epsfxsize=17.9cm
\centerline{\epsfbox{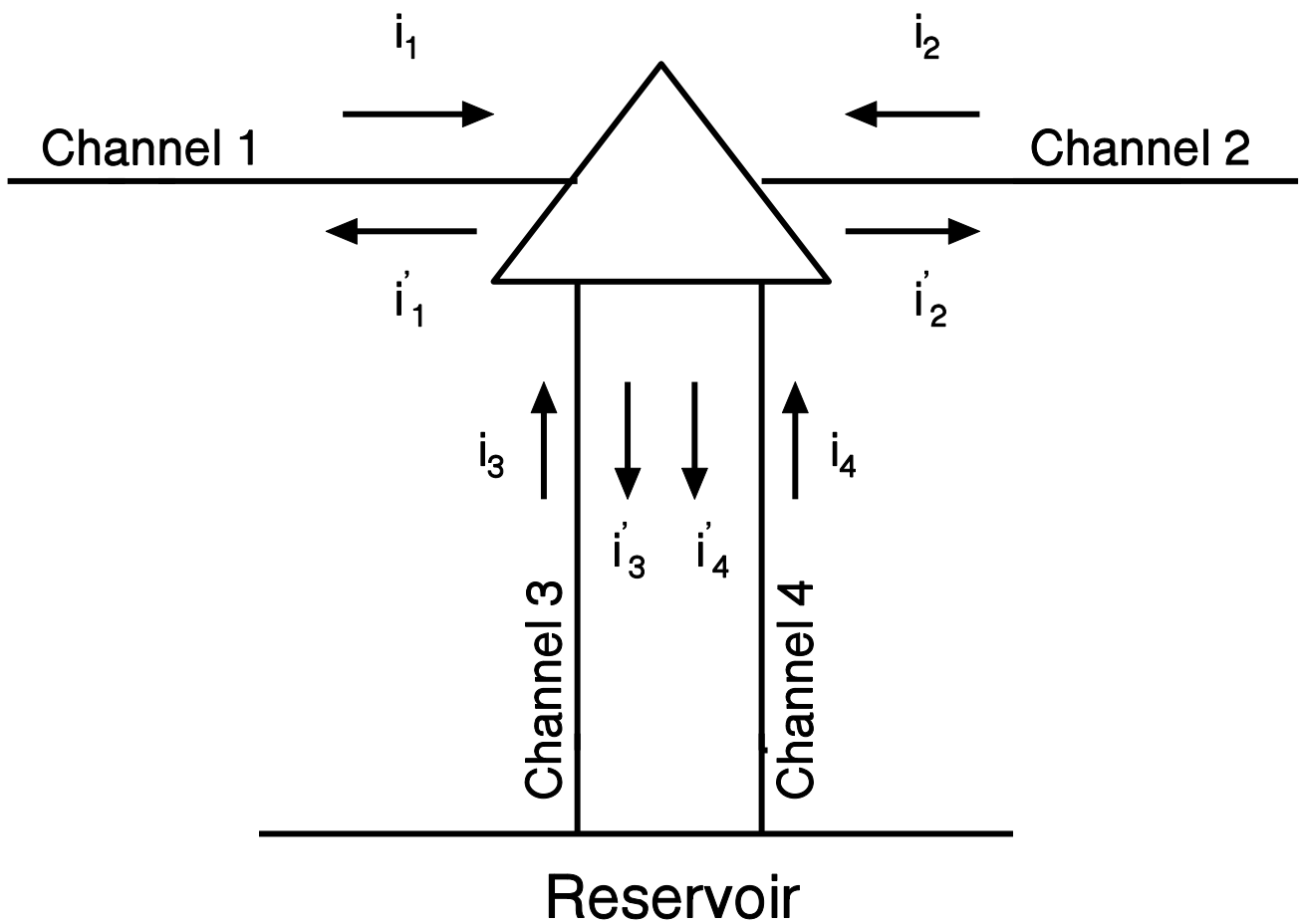}}
\caption{ Modeling "absorption" by fake channels:
Channels 1 and 2 are coupled through the  current leads to
two "fake" channels 3 and 4 which connect to a thermal reservoir.}
\label{figfake}
\end{figure}

The main idea is to simulate absorption by enlarging the
S-matrix to include some "fake", or "side" channels that remove
few  probability flux out of reckoning. 
Consider  scattering channels 1 and 2 connected through
current leads to two quantum "fake" channels 3 and 4 that carry
electrons to the reservoir, ( with chemical potential $\mu$) as
shown in the Fig.\ref{figfake} (for only one scatterer). This is a
phenomenological way of modeling absorption \cite{buttiker,maschke}.
Electrons entering into the channels 3 and 4 are absorbed regardless
of their phase and energy. The absorption is proportional to the
 strength of the "coupling parameter" $\epsilon$. 
A symmetric scattering matrix $S$ for such a
system can be written as :
%Let, us assume that the scatterer has general scattering channel
%and absorbing channel end with a phase randomizer which act as a blackbody.
%The scattering matrix can be taken as \cite{buttiker}, 
\begin{equation}
\left( \begin{array}{cccc}
            \alpha R  & \alpha T  &  0           & \beta       \\
            \alpha T  & \alpha R  & \beta       &  0          \\
             0        & \beta    & -\alpha R^*  & -\alpha T^*  \\
             \beta    &  0        & -\alpha T^*  & -\alpha R^*   
\end{array} \right) \quad ,
\label{meteq}
\end{equation}
where $R$ and $T$ are the reflection and the transmissions
amplitudes for the single scatterer , $\alpha =\sqrt{1-\epsilon}$
is the absorption coefficient and $\beta=\sqrt{\epsilon}$.\\ 
Now, the full scattering matrix is unitary 
for all positive real  $\epsilon < 1$. \\

But the sub-matrix connecting directly the channels 1 and 2 only
is not unitary. We will now explore this fact. We observe from the
above sub-matrix that for every scattering involving channels 1 and 2
only the reflection and the transmission amplitudes get multiplied by
$\alpha$ (the absorption parameter). Now keeping this fact in mind,
we will derive the Langevin equation for the reflection amplitude
$R(L)$ for $N$ scatterer (of length L) each characterized by a random
S-matrix of this type, with the same $\alpha$ value.

 Consider $ N $  random scatterers for a 1D sample of length $L$ 
with $(\equiv Na, a\equiv unit spacing)$ the reflection amplitude
$R(L)$, and $N+1$ scatterers of the sample length $L+\triangle L$,
with the reflection amplitude $R(L+\triangle L)$. That is, let us
start with the $N$ scatterers, and add one more scatterer to the
right to make up the N+1 scatterers. Now, we want to see the relation
between $R(L)$ and $R(L+\triangle L)$. That is, let there be a
delta-function potential scatterer between $L$ and $L+\triangle L$,
positioned at the point $L +\triangle L/2$, that can be considered as
the effective scatterer due to the extra added scatterers for length
$\triangle L$. 
For $k\triangle L\ll 1$, we can treat the extra added scatterer as an
effective delta potential $v_0(L) \delta (x-L- \triangle L/2)$
with $v_0(L) = V(L) \triangle L $.
(We consider the continuum limit, $a\rightarrow 0$,
 $N\rightarrow \infty$, $Na=L$, fixed ).

Now, for a plane-wave scattering problem for a 
delta function potential of strength $v_0$, which is at $x=0$
and has complex reflection and transmission amplitudes $r_0$ and $ t_0$
respectively, we have from the continuity condition for the wave function
and discontinuity condition for the derivative of the wave
function (which one gets by integrating Schr\"{o}dinger equation
across the delta function):\\ 
$ r_0 = \frac{v_0}{i 2 k - v_0} $ \\
and  $t_0 = r_0+1$. \\
Considering $ v_0 = v(L) \triangle L $ the smallness parameter, one gets
expressions up to first order in $\triangle L$ for $r_0$ and $t_0$:\\
$ r_0 = \frac{v_0 }{2ik}  $ and\\
$ t_0 = 1 + \frac{v_0 }{2ik} $, where we have taken 
$\hbar^2/2m=1$.\\

Now to introduce absorption we  will write:\\
 $  r \rightarrow r \alpha = r_0(1-\epsilon)^{1/2}  $, \\ 
 $  t \rightarrow t \alpha = t_0(1-\epsilon)^{1/2}  $. \\

This means that for every scattering the reflection and transmission
amplitudes get modulated by a factor of ${\alpha}$.

Now consider a plane wave incident on the right side of the
sample of length $L+\triangle L$. Summing all the processes of
direct and multiple reflections and transmissions, on the right
side of the sample of length $L$, with the effective delta
potential at $L+\triangle L/2$, one gets, 
\begin{eqnarray}
 R(L+\triangle L) &=& r e^{ik\triangle L} \nonumber \\
    &+& e^{ik\triangle L/2} t e^{ik\triangle L/2}  R(L) 
           e^{ik\triangle L/2} t e^{ik\triangle L/2} \nonumber \\
       &+& ....... \,.
\label{2.24}
\end{eqnarray}
 Summing the above  geometric series, substituting the values of $r$ and $t$, 
and taking the continuum limit for L , one gets from Eq.\ref{2.24},
the Langevin equation:
\begin{equation}
 \frac{d R(L)}{d L} \, = \, - \alpha R(L) +  
2ik R(L)\, + \,i \frac{k}{2} (\eta(L))(1+R(L))^2 \, ,
\end{equation}
with the initial condition $R(L)=0$ for $L=0$, and $\alpha $ is the
absorbing parameter. 
This is not quite the same as the Eq(4.4) obtained by
introducing an imaginary potential. It turns out, however, this Langevin
equation gives same results in the regime of weak disorder but
differs qualitatively in the regime of strong disorder.

\vspace{.5cm}
\centerline{\bf Acknowledgements}
Most part of this work was done at the Physics Department,
Indian Institute of Science, Bangalore.
I gratefully acknowledge N. Kumar for many stimulating discussions
and pointing out many pictures of physical interest and several
suggestions. I also thank to the Council for Scientific and Industrial
Research (CSIR), India for financial help.

%\newpage

%  
.
%\twocolumn[\hsize\textwidth\columnwidth\hsize\csname@twocolumnfalse\endcsname
\vspace{.21cm}
\newpage
\end{document}